# Anapedesis:

# Implications and Applications of Bio-Structural Robustness

Nicanor I. Moldovan*

*Davis Heart and Lung Research Institute,*
*Departments of Internal Medicine/Cardiology, Ophthalmology and Biomedical Engineering*
*Ohio State University, Columbus, OH 43210, USA*

Here we develop an approach to bio-structural robustness integrated with structure-function relationship in a unified conceptual and methodological framework, and envision its study using adequate computational and experimental methods. To distinguish this structural robustness from the abstract organizational robustness of systems, we call it **anapedesis,** and define it as the *scale-independent property of biological objects, from biomolecules to organisms, to deform and recover while minimizing and/or repairing the damage produced by stretch*. We propose to study the consequences of deformation of biological objects closer to their structural and/or functional failure than previously considered relevant. We show that structural robustness is present as a basic principle in many facets of biomedicine: many pathological conditions may derive from the failure of molecules, cells and their higher-order assemblies to maintain robustness against deformation. Furthermore, structural robustness could have been the key selective criterion during pre-biotic evolution and afterwards, and its universality can be demonstrated by modeling using genetic algorithms. Thus, the specific investigation of bio-structural robustness as anapedesis could help the solving of fundamental problems of biology and medicine.

## 1. Introduction

From the structure-function paradigm of modern biology, it follows that for maintaining and optimally manifesting their functions, biological structures should be essentially elastic, i.e. able to promptly recover their structure after the ubiquitous deformations that are co-substantial with life. Given the natural plasticity of polymers, biological materials thus need a compensatory property to minimize it, providing them with the desired structural robustness[3].

So far, the emphasis was placed on the organizational, systems-related robustness[3;4]. However, mechanical robustness of biomaterials was not systematically addressed yet. This notion received attention only as 'stability' of structural pattern of biomolecules[5], being related to their evolvability[6], as shown by modeling[7].

Biological structures at all levels of organization are endowed with a built-in design allowing protection of their structural integrity. These are also equipped with the sensing of incoming structural threats, and with mechanisms for restoring or removal (when not possible otherwise) of the damaged components. In this category are included functions of nervous system (touch, pain), cellular mechano-sensory mechanisms, or secretion of *alarmins*[1], a class of specialized molecules that signal to cells with repairing function. Accumulation of structurally deficient proteins or nucleic acids in cells leads to the activation of a set of "*alarm genes*"[2] that triggers removal of the affected cells, when repairing becomes inefficient.

I propose that the components of living beings, at all levels of organization, embody a so far poorly conceptualized property, consisting in their *ability to actively and complexly deform, minimizing the damage produced by stretch, and to recover and/or to repair structural damage when it inevitably occurs*. To address this property, I propose '*anapedesis'*, a Greek word for 'resilience', 'recoil', or 'bouncing back' (a notion used by Aristotle to describe pulsations of heart, in *De vita et morte,* Cap. II, *De Respiratio*, Engl. tr. J. E. Beake, Oxford, 1908, tom III, page 553).

What differentiate anapedesis from common visco-elastic deformation, and cannot be derived from materials science without explicit consideration are: (a) structural design minimizing the consequences of stretch, (b) biological processes concurring to non-traumatic deformation; and (c) active post-deformation recovery and/or repair. Sometimes the deformation, recovery and/or repair are autogenous, but in most cases are assisted by cooperating structures. In general, anapedesis is manifested each time the environment imposes a stretching upon biological objects (e.g. during passing through conduits with size smaller than its cross section), or as effect of shear stress.

## 2. Cellular anapedesis.

During deformation of cells, subjacent processes concur to prepare it, assist with shape recovery and minimize and/or repair the damage. More than material deformability, this represents a particular *cell behavior,* which I refer to as 'cellular anapedesis'. This is involved during any constrained movement, such as the *longitudinal* squeezing of cells in microvessels, or the transversal migration of leukocytes during *diapedesis* across the vessel wall[8]. In leukocytes this includes dynamic regulation of cell stiffness[9], with initial softening of cytoplasm and paradoxical increase of volume consecutive to water influx[10]. Post-stretching actin polymerization temporarily stabilizes the cells in an elongated form[11], likely preparing those already localized in a microvessel for the next squeezing. During deformation, a pre-existing excess plasma membrane reserve, stored in surface folding, is utilized to accommodate the increase in surface/volume ratio[12].

In a variety of cells and conditions plasma membrane often breaks, but is promptly repaired[13]. This repairing is based on a population of sub-membrane vesicles[14], ready to fuse with plasmalemma[15]. During this process extracellular molecules could enter the cells (e.g. albumin in cardiomyocytes of hypertensive animals[16], or in aortic endothelial cells[17]). Alternatively, signaling molecules could escape from the cells in physiologically-meaningful amounts[18]. Therefore, deformation-induced 'failure' of plasma membrane integrity could be beneficial, and thus allowed yet tightly controlled, for instance during the unconventional secretion of a variety of proteins lacking a signal sequence[19;20]. Among these, bFGF is prototypical and thought to be released from cells by either sub-lethal damage[19;20], or by yet another mechanism involving a trans-membrane release with direct contribution of the *external*, glycocalix-associated proteoglycans[21,2].

When the limiting objects amongst the cells perform anapedesis is are biological, they all would cooperate for successful passage: the conduit will enlarge appropriately but remain in close apposition. This property is known as *compliance,* described in cardiovascular system[22], lungs etc. A key regulator of deformations of all involved structures nitric oxide (NO), appreciated mostly for inducing relaxation of vascular smooth muscle cells. However, in the same paradigm should be included its effects in reducing endothelial stiffness[23], or concentration-dependently regulating the deformability of erythrocytes[24], leukocytes[25], and progenitor cells[26]. Therefore, NO qualifies as a complex anapedetic regulator. This highlights the possible involvement of NO-regulated anapedesis in most instances of translocation of inflammatory cells, as during pathogenesis of atherosclerosis, hypertension, sepsis, autoimmune disorders, etc.

## 3. Organelle anapedesis.

Cell deformations are propagated downwards to organelles. The tensegrity model of cytoplasm[27] suggests that the cell is normally in a pre-stressed state (possibly to favor a quasi-elastic deformation, and a more efficient shape recovery). However, less is known about this behavior in organelles. Biomechanical properties of *nuclei* were more extensively studied[28], because the expression of many genes is mechanically sensitive[29], probably as effect of a 'top-down' propagation of mechanical stress in a molecularly crowded space[30]. This property is developmentally regulated, the nuclei being more deformable during the primitive stages, when cellular translocations are more active, and becoming stiffer with differentiation[31].

*Mitochondria* have very active autonomous intracellular translocations[32]. Since they often display an elongated shape[32], this might be an anapedetic property, avoiding large changes in the surface/volume ratio consecutive to deformation.

*email: nicanor.moldovan@osumc.edu

This is due to the fixed external mitochondrial membrane area, but the presence of a repairing mechanism was not investigated so far. A component of 'permeability transition pore' seems to be the failure of inner mitochondrial membrane integrity, consecutive to osmotic imbalance[33].

Mitochondria are exchanged between adult and stem/progenitor cells, in a process of metabolic 'rejuvenation'[34]. This exchange takes place via sub-micron wide 'nanotubes', thin cytoplasmic extensions, in contact with the target cells, within which mitochondria extensively squeeze[35]. If and how this happens without organelle damage, is not known.

### 4. Molecular anapedesis.

Anapedesis applied to bio-molecules is the expression of robustness of their deformations, in relation to biological functions. This knowledge is required for a thorough understanding of structural robustness of biomolecules in native environments, as well as in the artificial circumstances created during single-molecule manipulation for nano-technological applications[36;37].

The ample changes in configurations of macromolecules or macromolecular assemblies, either programmed (e.g. disassembly of chromosomes, reversible unwinding of DNA, etc), or spontaneous (such as intermolecular structure propagation among prions and other amyloid-forming proteins[38]), are likely to reflect the anapedetic behavior of molecules. Compactness of nucleic acids in chromosomes[39] and in viruses[40], as well as their ability to safely and efficiently re-pack after unwinding, are also excellent illustrations of molecular anapedesis.

Cell-scale deformations are transmitted down to molecules via a poorly understood 'mechano-sensing' mechanism. To this category belongs, for example, the impact of shear stress or shape changes on gene expression and cytokine secretion[29]. Recently, it was shown that mechanical deformations can be propagated from macro-scale to molecular level in polymers, as reflected by a change in their engineered enzymatic reactivity[41].

Unexplored anapedetic rules of organization are embodied in the structure of molecular channels able to maintain their selectivity under mechanical stress, trans-membrane transport of molecules and their insertion into the endoplasmic reticulum membrane during biosynthesis, nuclear and mitochondrial transport, or in the organization of cytoskeletal molecules. These complex, potentially-reversible deformations are often assisted by other molecules (chaperones), which can also perform not only refolding activities, but also aminoacid-level repairing[42].

In a recently proposed model of cytoplasm[44], biomechanics of biomolecules cooperatively generate the non-linear glassy material parameters of the cytoplasm[43] and of nuclear matrix[30]. These properties were suggested to derive from molecular 'jostling'[45], in which water serves both as lubricant and as a bond, therefore as a structural component. Water's chemical activity determines the status of actin polymerization[46], and thus cytoplasm stiffness[47], making anapedesis ultimately dependent on water's properties.

Filtration of plasma molecules in the renal glomerulus is based on their *deformability,* rather than on their molecular weight and/or surface charge[48-50]. This implies that the kidneys use the proper deformability (i.e. molecular anapedesis) to sort and filter out damaged molecules, which having lost their proper assembling, are unable to perform their functions.

### 4. Applications

Essential information about cellular anapedesis could be obtained from how much reversible, non-lethal deformation various cell types could sustain, as well as from the reversibility of the mechanical cell damage in general. The study of post-traumatic survival of cells and organisms (as well as the detailed conditions of damage repair), could find immediate translation in resuscitation medicine.

To quantify cellular anapedesis, I defined a set of adimensional coefficients obtained by comparison to internal standards, based on: (1) accumulation of plasma membrane impermeable probes (such as propidium iodide) into cells during controlled deformations[51]: (2) cell volume; (3) plasma membrane area; (4) total and cortical (plasma membrane-stabilizing) actin polymerization; (5) generation of intracellular signals influencing actin polymerization, such as reactive oxygen species (ROS)[52] and/or NO[26]. Because these variables can be concurrently assessed using fluorescently labeled molecules, the proposed anapedesis assay is based on a flow cytometry platform.

Anapedesis might control the fate of old and deficient circulatory cells, or of non-circulatory cells accidentally (e.g. tumor cells) or purposefully (such as those administered for cell therapy) present in the bloodstream. Interestingly, the earliest changes in drug-induced apoptosis of leukemia cells, even before caspases activation, is a two orders of magnitude increase in their stiffness[53]. From this, and for avoiding the development of a strong pro-coagulant cell surface specific for fully apoptotic cells[54], we anticipate that the liver and spleen precociously retain dysfunctional circulating cells, mainly based on their modified deformability. Consecutively, the optimization of cellular anapedesis to reduce non-specific organ retention, would expand the lifetime in circulation and the accessibly and survival of cells in the heart.

*email: nicanor.moldovan@osumc.edu

*A. Improving the efficiency of cell therapy.* Currently, the efficiency of *in vivo* administration of stem/progenitor cells targeted to vital organs is very low, because the infused cells transiently accumulate in the lungs, then in liver, spleen, and bones[61]. Poor post-engraftment viability is the key factor, because the overexpression of the survival factor Akt dramatically improves the engraftment of these cells[62].

We propose that a major limitation of successful retention of both freshly isolated stem/progenitor cells, or of those *in vitro* amplified and then injected into animals, is the absence of properties that make the leukocytes able to efficiently and safely travel through narrow capillaries, i.e. a proper anapedetic behavior. For example, mesenchymal stem cells in circulation lack the osmotic properties of erythrocytes, making them insensitive to osmotic stress[63], a condition for proper anapedetic behavior[10]. Similarly to tumor cells, known to be very sensitive to stretch during diapedesis[12], the therapeutic stem cells might be at the same risk, essentially because they do not have the anapedetic properties of leukocytes.

Leukocytes are temporarily retained in lung microvasculature, initially based on their cytoplasmic stiffness, and only later engaging the surface adhesion molecules[47]. In order to avoid the consequences of this retention, the lungs are endowed with a parallel organization of microvessels, which minimizes the impact on erythrocyte flow of capillary obstruction. How cell retention proceeds in other organs, such as muscles, is less clear. We think this involves the notorious 'collateral flow' system, consisting in microvascular branches of larger diameter, which would allow leukocytes to bypass the capillary system.

*B. Fate of bone marrow-derived progenitor cells during extracorporeal circulation.* Failure of cellular anapedesis, mostly that of erythrocytes, underlies the alteration of blood quality during the use of either extra-corporeal ventricular assist devices[55], or during cardio-pulmonary bypass[56]. This is due to direct cell damage inflicted by high shear stress and/or the interaction with non-compliant interfaces. Less is known about the impact of extracorporeal circulation and of filter-based depletion on leukocytes (making the patients on extravascular circulation susceptible to immune disorders[57]). Even less information is available about the fate in these conditions of circulating progenitor cells, involved in the maintenance and repair of endothelial lining and various organs[58] and the heart[59].

Therefore, we propose to study anapedesis of shear-stressed leukocytes and progenitor cells in suspension, or collected from extra-corporeal circulation devices, and to improve it by targeted interventions. Erythrocytes treated with NO-generating drugs, or the blood from patients taking statins, are more resistant to this damage, apparently consecutive to erythrocytes membrane stabilization[56]. We further anticipate that rational design of cell deformability by using cytoskeleton targeting drugs, combined with plasma membrane stabilizers such as poloxamer 188[60], could improve the quality of circulating leukocytes and progenitor cells.

*C. Effects of statins on cellular biomechanics.* A fundamental property of neoplasms is an *amplified* robustness at various levels of organization, apparently covering a hidden vulnerability[71]. Finding these Achilles' heels in tumor robustness is now a high priority in cancer treatment research[71]. We suggest that a decrease in biomechanical robustness could be a key target for novel anti-tumor strategies. In cells with disturbed anapedetic response, not only their plasma membrane, but also the whole cytoplasm would easier break down. Indeed, tumor cells cannot withstand substantial elongation, as shown recently *in vitro*[12], and they often fragment when entering from the extravascular space into a blood vessel (reverse diapedesis)[72], or in the heart[73;74], possibly due to their considerably reduced stiffness[75].

Metastases usually occur in lungs, liver or bones, the organs known for non-specific cell retention, although in a very small proportion of initial insemination[76]. In these organs, tumor cells survival is apparently dependent on deformation-induced cell damage[77]. We expect that statins might influence survival and spreading of tumors by modifying their anapedetic, i.e. cytoskeleton- and plasma membrane-dependent properties, because they inhibit prenylation of RhoA-class molecules, interfering with their functions (such as Rac1-dependent actin polymerization[67] and ROS signaling[68]). Indeed, the anti-hypercholesterolemic treatment with statins has as indirect effect the change, often in better[69] but sometimes in worse[70], of the course of several cancers.

In muscles, stretch-induced plasma membrane repair proceeds via a dystrophin- and dyspherlin[57]-dependent, calcium-mediated[58] fusion of sub-membrane vesicles. Cortical F-actin network, otherwise needed for providing the cells with mechanical stability, interferes with, and controls, this repairing process[59;60]. We propose that statin-induced myopathy could be the consequence of inhibition of prost-translational prenylation of actin-regulating Rho-class molecules[52], and thus of muscle cells plasma membrane repair.

Rupture threshold of cytoplasm might also be sensitive to statin action. This is suggested by the fact that release of platelets into the circulation from megakaryocytes is a form of shear-stress dependent physiological cell fragmatation[66] and one of the side effects of statin treatment is an occasionally substantial, and probably more often sub-clinical, thrombocytopenia[64;65]. Therefore, the approach proposed here could bring a needed clarification in this field, and highlight an untapped therapeutic potential of statins.

*D. Redox-dependent molecular biomechanics.* Our interest in molecular anapedesis will focus on *reversibility* of molecular deformations, and in particular, on how biochemical environment facilitates or prevents this reversibility. For molecular anapedesis, redox biochemistry is of particular relevance. After mechanical deformation, key cytoskeletal proteins (e.g. myosin in the substrate-attached mesenchymal stem cells[79]), as well as several other proteins[80;81], expose reactive cysteines, otherwise buried in protected pockets. If redox balance of the cell is affected, then formation of either intra or inter-molecular disulphide bridges will be changed, altering not only the functionality of these molecules, but also their biomechanical properties. These in turn could collectively determine changes in the cytoskeleton-dependent properties of cells, such as motility, deformability or squeezing behavior[82]. Alternatively, in tissues under combined biomechanical and redox stress, such as in post-ischemic hearts, activity of antioxidant enzymes (e.g. thioredoxins) could be markedly affected, due to their molecular biomechanical sensitivity[81].

In addition, the deleterious effects of oxidative stress on cells (e.g. nitration of F-actin by peroxinitrite in erythrocytes[83] and leukocytes[84]), or of hyperglycemia (via NO-containing adducts, or AGEs, etc), might act by impairing the recovery from deformation of individual proteins, and in addition to other functional consequences, influencing the emergent mesoscale cytoplasm biomechanics[43].

*E. Modeling anapedesis.* Wound healing and the stem cells-based repair in adult tissues, such as satellite muscle-specific stem cell, is the organ-level expression of anapedetic recovery. Using first principles and a genetic algorithm, it was recently demonstrated that structures endowed with the ability to self-repair can be evolved *in silico*, showing the emergence of a subpopulation of structural units that retain indefinite proliferative capacity, analogous to stem cells[85]. Wound healing and the stem cells-based repair in adult tissues, such as satellite muscle-specific stem cell, can be conceived as the organ-level expression of anapedetic recovery. This shows that it is possible to demonstrate the emergence of anapedetic behavior at all scales, by abstract representation of biological objects.

Because molecular anapedesis might have been the ground of pre-biotic selection, even before 'survival' had a meaning, and long afterwards, I propose to test if anapedesis is an evolution-derived property of deformable structures, and verify whether anapedesis, as a fundamental property of biological organization, could be directly derived from basic principles by computer modeling.

The ontological consistence of the *individual* (in fact, of the *self*) justifies an agent based modeling (ABM) approach for biology. Reversibly, the appropriateness of artificial intelligence-derived models to describe basic principles of biological organization argues that the individual is the main biological character, the object of selection and of evolution, and not the population or the (quasi)species to which it belongs.

**5. Implications for pre-biotic evolution and biogenesis.**

*A. Structure-function.* I propose that in the deepest sense, 'anapedetic' is synonymous with 'biological', and could be used as its practical definition. Assays based on anapedesis could become essential tools of searching for life in alien environments.

*email: nicanor.moldovan@osumc.edu

Biological structures essentially represent spatial distribution of molecules and of sub-molecular elements; biological functions are thought to derive from these structures (based on the non-equilibrium character, derived from their chemical reactivity). Thus, structural biology is based on the assumption that *function derives from structure*. However, what is a 'biological' structure, how function emerges from structure, and how to accommodate the numerous variants and exceptions to this fundamental law (including a *'function-driven adaptation'* we described in the stem cells world, or the *exaptation* to a new function of pre-existing structures, etc), is poorly understood.

I propose that emergence of complex behaviors from subjacent structures is the expression at a higher scale of organization of this unifying principle, namely of anapedesis. *Since not all structures are functional, I make the conjecture that functional are only the anapedetic structures, i.e. those embodying a design (obtained by natural selection) which provides them a 'self' quality, susceptible to structural recovery and/or repair.*

*B. Individual.* Living matter is organized as a collection of individuals, which maintain both their individuality (as separateness), and their integrity (internal connectivity). Structural integrity is a fundamental property of living beings. It also defines what an **individual** is: a portion of living matter that maintains its structural integrity. Sometimes the individuals trade off their individuality for association in larger structures, which starts to behave like a new unity, i.e. as a new individual. Alternatively, cell division, which generates two new individuals, is an incorporated, self-imposed anapedesis failure, i.e cytoplasm 'breaking'.

The definition of the *'biological individual'* is very complex, and far from understood (see the difficulty to define it in non-animal settings); it is tempting to speculate that what gives the biological individual ontological consistence is inscribed in its structure from genesis.

*C. Self.* Structural integrity is one key factor of individual stability, possibly its definition; and thus, anapedesis becomes central to the concept of individual. Conservation of structure by robust design and active restoration defines the biological meaning of 'self', of fundamental importance for biology. Combined with the concept of 'biological individual', anapedesis is at the core of biological 'being' endowed with an autonomous self.

Anapedetic-type structure and dynamics at molecular and organelle level becomes *behavior* at cellular and organism level. The link is anapedesis, as the active mechanism of (self-)repair. Selfness would then represent the *maintenance* of a structure (upon spatial-temporal transformation), combined with its *repair*.

Evolution of biological structures leading to the acquisition of 'function' must have been in close relationship with the improvement of their robustness. In fact, *I suggest that this was the selection criterion operating in the pre-biological world,* before 'survival', 'proliferation', and/or heredity became possible.

Thus anapedesis offers an alternative, complementary and novel, more general selection criterion, operable before biogenesis. This was shown *in silico* using cellular automata and genetic algorithms[85]. In this model, although abstract genes warrant the stability and continuity of the structures, the subject of evolution is *not* these 'selfish genes' spreading via population multiplication, but the scale-independent individual prototype (approximately represented by any concrete member of the group, i,e of quasi-species), self-selecting and thus and self-perfecting for, and by, an ever increasing robustness.

In order for repairing process to take place, it is necessary that the structure to be dynamic (i.e. the components in constant *turn-over*), rather than static. In this case, the repairing is simply a process of *replacement* of the parts with new, undamaged sub-ensembles which would come in place. Another feature of the repairing activity is the need for inter-subunit cooperation (inter-molecular, with contribution of enzymes, inter-organelle, inter-cellular etc), the stochastically distributed traits of the sub-ensembles facilitating the availability of those with the appropriate properties.

These principles have relevance for the emerging synthetic biology, where the substitution of elements from biological world with artificial replacements requires the observation of basic rules of assembly and functioning of the native, biological counterparts.

*D. Biogenesis.* Gravity increases the density of matter. In a dense corner of the Universe (such as a planet), no movement is free from constraints. The common constraint to movement of an object is friction with the environment. Another is interaction with other objects, such as collision. Both impose a shape adjustment to the moving object, as a first example of 'selective pressure'.

*email: nicanor.moldovan@osumc.edu

In pre-biotic and unicellular stage, the main mechanical damaging force was fluid shear stress; later, in multicellular stage, to these translocations or percolations of cells within the organism produced another type of biomechanical tension, derived from their interaction.

The most recent models of the cytoplasm are based on the assumption (and realization) of an extremely crowded molecular space. This imposes severe limitations to movement, where interaction of all neighbors is reciprocally conditioned and constrained in biomechanical terms. Also, it imposes substantial selective pressure on the shapes and/or deformations these objects could undergo. If the deformations are irreversible and lead to loss of 'function', the system would functionally collapse.

*In fact, it is conceivable that 'function' is exactly this ability to self-sustain and - by integration - to sustain the 'self' of other molecules.* Over time, this leads to 'selection' of structural transformations compatible with 'molecular survival', i. e. maintenance of structural integrity. It is conceivable that the whole metabolism is a sophisticated manifestation of the (self-)repairing process of molecules, namely of anapedesis.

It is significant that the anapedetic, (self)repairing property of molecules relies not only on the ability to autonomously perform this activity, but also - if not mostly - on inter-molecular cooperation via chaperones, 'enzymes', and alike. The distinction is in fact almost arbitrary. Therefore, I propose that the 'auto-enzymes' are ontologically more primitive than those where this activity is dedicated to 'others'.

Auto-catalysis was probably the basic, and most ancient manifestation of life. Remarkably, the most vestigial biological function known, RNA splicing, is a self-repairing process which might have been at the core of the pre-biotic "RNA world". Deep down, the distinction between the self-repairing activities and the myriad post-translational modifications with various consequences (functions), related either to supra-molecular complexes assembly or to 'signaling', becomes superfluous.

The earliest forms of cell division possibly were the result of a 'regulated' breaking of the pre-cellular 'organoid', maybe as effect of shear stress in the aquatic milieu (anapedesis failure). This would have been later incorporated in the standard cell multiplication process (a more contemporary example is the generation of platelets from megacaryocytes). Morphologically, the cleavage furrow looks like a 'tube' forcing the cytoplasm to squeeze.

Organism-level anapedesis was (and is) essential for movement of living things in aqueous milieu: reduction of friction requires fusiform shape, which leads to formation of 'fluid tubes' which are tubes in which an object moves. Extracellular matrix development proceeded in this selective environment.

Fragility of organisms was from the beginning a matter of evolutionary pressure and/or selection. The Cambrian explosion of trilobites might be related to the development of the external shell, allowing the maintenance of integrity in face of the shearing/breaking forces of the environment, and thus the expansion in biomechanical niches by then inaccessible. Another similar example is the eggshell, as 'anapedetic' shield protecting the embryo during the critical, mechanically sensitive phases of individual development.

The principle of anapedesis allows the understanding of life in a new way (biophysical and/or biomechanical, rather than biochemical/metabolic as in the current definitions). Therefore, it comes in *complementation* to the current metabolic, reproductive and informational aspects of living; beyond the biochemical features, it could help us detect life elsewhere in the Universe.

**6. Anthropogenesis.**

At a higher level or organization, the body plan of all mammals needs to be compatible with (and is anatomically designed to perform) many deformations. For example, birth canal needs to be maintained at minimal expansion. This requires that fetal skeleton should be at birth in a state of optimal deformability (including the skull), thus putting severe limits for the duration of gestation, body weight and the level of maturation at birth.

A most remarkable situation pertains to our own species: because the increased size of the brain, the birth of human babies became proportionally complicated and risky for both mother and progeny. It has been suggested that evolutionarily, postural repositioning driven by increasingly bipedal movement of the pelvic bones has restrained the birth canal, and thus the already increased primate brain size. This apparently selected for a more deformable skull, leading to a delay in neural maturation and increased familial social dependency, which is a biological pre-requisite for cultural development. Thus, a key human feature might have been the product of anapedesis.

*email: nicanor.moldovan@osumc.edu

*email: nicanor.moldovan@osumc.edu

*email: nicanor.moldovan@osumc.edu*